\documentstyle[11pt,epsfig]{article}
\begin{document}
\title{\Large\bf  The $\eta(2225)$ observed by the BES Collaboration}

\author{\small De-Min Li \footnote{E-mail: lidm@zzu.edu.cn},~~Bing Ma\\
\small  {\sl Department of Physics, Zhengzhou University,
Zhengzhou, Henan 450052, People's Republic of China}}
\date{\today}
\maketitle
\vspace{0.5cm}

\begin{abstract}

In the framework of the $^3P_0$ meson decay model, the strong
decays of the $3\,^1S_0$ and $4\,^1S_0$ $s\bar{s}$ states are
investigated. It is found that in the presence of the initial
state mass being $2.24$ GeV, the total widths of the $3\,^1S_0$
and $4\,^1S_0$ $s\bar{s}$ states are about $438$ MeV and $125$
MeV, respectively. Also, when the initial state mass varies from
2220 to 2400 MeV, the total width of the $4\,^1S_0$ $s\bar{s}$
state varies from about 100 to 132 MeV, while the total width of
the $3\,^1S_0$ $s\bar{s}$ state varies from about 400 to 594 MeV.
A comparison of the predicted widths and the experimental result
of $(0.19\pm 0.03^{+0.04}_{-0.06})$ GeV, the width of the
$\eta(2225)$ with a mass of $(2.24^{+0.03+0.03}_{-0.02-0.02})$ GeV
recently observed by the BES Collaboration in the radiative decay
$J/\psi\rightarrow\gamma\phi\phi\rightarrow\gamma
K^+K^-K^0_SK^0_L$, suggests that it would be very difficult to
identify the $\eta(2225)$ as the $3\,^1S_0$ $s\bar{s}$ state, and
the $\eta(2225)$ seams a good candidate for the $4\,^1S_0$
$s\bar{s}$ state.

\end{abstract}

\vspace{0.5cm}
 {\bf Key words:} mesons, $^3P_0$ model

 {\bf PACS numbers:}14.40. Cs, 12.39.Jh

\newpage

\baselineskip 24pt

\section{Introduction}
\indent \vspace*{-1cm}

Experimentally, a low-mass enhancement in $J/\psi$ radiative
decays $J/\psi\rightarrow \gamma\phi\phi$ at $2.25$ GeV with a
clear pseudoscalar assignment was first reported by the DM2
Collaboration\cite{e1}. Subsequently, the DM2
Collaboration\cite{e2} and the MARK III Collaboration\cite{e3}
gave the evidence of a resonant $\phi\phi$ production around $2.2$
GeV, preferably pseudoscalar, also in $J/\psi\rightarrow
\gamma\phi\phi$.  A fit to the $\phi\phi$ invariant-mass spectrum
gave a mass of $(2230\pm 25\pm 15)$ MeV and a width of
$(150^{+300}_{-60}\pm 60)$ MeV\cite{e3}. An angular analysis of
the $\phi\phi$ signal found it to be consistent with a $0^{-+}$
[$\eta(2225)]$ assignment. The nature of the $\eta(2225)$ is
unclear. Possibilities of the nature of the $\eta(2225)$ include
the second and third radial excitations of the pseudoscalar meson
$\eta^\prime$, hybrid, glueball or multiquark state. However, the
large uncertainty of the width of the $\eta(2225)$ leads to that
the theoretical interpretations perhaps remain open.

Recently, based on the $5.8\times 10^7$ $J/\psi$ events collected
in the BESII detector, the radiative decay
$J/\psi\rightarrow\gamma\phi\phi\rightarrow\gamma
K^+K^-K^0_SK^0_L$ was analyzed by the BES Collaboration, and a
near-threshold enhancement was found in the $\phi\phi$ invariant
mass distribution at $2.24$ GeV with a statistical significance
larger than $10~\sigma$. A partial wave analysis shows that this
structure is dominated by a $0^{-+}$ [$\eta(2225)$] with a mass of
$(2.24^{+0.03+0.03}_{-0.02-0.02})$ GeV and a width of $(0.19\pm
0.03^{+0.04}_{-0.06})$ GeV, and the production branching fraction
is $\mbox{Br}(J/\psi\rightarrow
\gamma\eta(2225))\mbox{Br}(\eta(2225)\rightarrow \phi\phi)=(4.4\pm
0.04\pm 0.8)\times 10^{-4 }$\cite{BES08}. The improved
measurements of the $\eta(2225)$ performed by the BES
Collaboration maybe open a window for revealing the nature of the
$\eta(2225)$.

It is very important to exhaust possible conventional $q\bar{q}$
description of the $\eta(2225)$ before resorting to more exotic
interpretations such as hybrid, glueball or multiquark state as
mentioned above. In the present work, we shall focus on the
possibility of the $\eta(2225)$ being the ordinary pseudoscalar
$q\bar{q}$ state. From PDG2006\cite{pdg2006}, the $1\,^1S_0$ meson
nonet ($\pi$, $\eta$, $\eta^\prime$ and $K$) as well as the
$2\,^1S_0$ members [$\pi(1300)$, $\eta(1295)$ and $\eta(1475)$]
have been well established. In our previous work\cite{eta1835}, we
suggested that the $\pi(1800)$ and $K(1830)$, together with the
$X(1835)$ and $\eta(1760)$ observed by the BES
Collaboration\cite{x1835,BES17602}, constitute the $3\,^1S_0$
meson nonet. Theoretically, both the second\cite{cqm} and
third\cite{regge} radial excitations of the $\eta^\prime$ are
predicted to lie in the mass range of the $\eta(2225)$. The main
purpose of this work is to evaluate the widths of the $3\,^1S_0$
and $4\,^1S_0$ $s\bar{s}$ states in the $^3P_0$ meson decay model,
and then check which of these two pictures can reasonably account
for the total width of the $\eta(2225)$.

The organization of this paper is as follows. In section 2, the
brief review of the $^3P_0$ decay model is given (For the detailed
review see {\sl e.g.}
Refs.\cite{3p0rev1,3p0rev2,3p0rev3,3p0rev4}.) In section 3, the
decay widths of the $3\,^1S_0$ and $4\,^1S_0$ $s\bar{s}$ states
are presented, and the summary and conclusion are given in section
4.

\section{ The $^3P_0$ meson decay model}
\indent \vspace*{-1cm}

 The $^3P_0$ decay model, also known as the quark-pair creation
model, was originally introduced by Micu\cite{micu} and further
developed by Le Yaouanc et al.\cite{3p0rev1}. The $^3P_0$ decay
model has been widely used to evaluate the strong decays of
hadrons\cite{3p00,3p0y,3p0x,3p0x1,3p0x2,3p01,3p02,3p03,quarkmass,3p04},
since it gives a good description of many of the observed decay
amplitudes and partial widths of the hadrons. The main assumption
of the $^3P_0$ decay model is that strong decays take place via
the creation of a $^3P_0$ quark-antiquark pair from the vacuum.
The new produced quark-antiquark pair, together with the
$q\bar{q}$ within the initial meson regroups into two outgoing
mesons in all possible quark rearrangement ways, which corresponds
to the two decay diagrams as shown in Fig.1 for the meson decay
process $A\rightarrow B+C$.

\begin{figure}[hbt]
\begin{center}
\epsfig{file=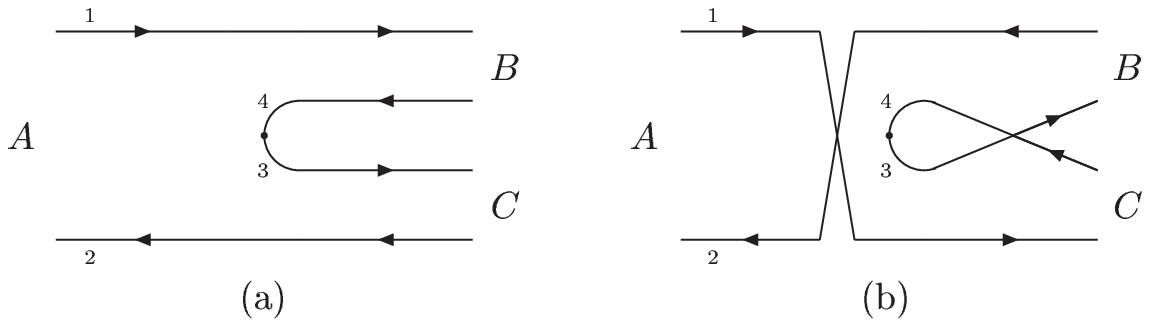,width=12.0cm, clip=}
\vspace*{0.5cm}\vspace*{-1cm}
 \caption{\small The two possible diagrams contributing to $A\rightarrow B+C$ in the $^3P_0$
 model.}
\end{center}
\end{figure}

The transition operator $T$ of the decay $A\rightarrow BC$ in the
$^3P_0$ model is given by
\begin{eqnarray}
T=-3\gamma\sum_m\langle 1m1-m|00\rangle\int
d^3\vec{p}_3d^3\vec{p}_4\delta^3(\vec{p}_3+\vec{p}_4){\cal{Y}}^m_1(\frac{\vec{p}_3-\vec{p}_4}{2})\chi^{34}_{1-m}\phi^{34}_0\omega^{34}_0b^\dagger_3(\vec{p}_3)d^\dagger_4(\vec{p}_4),
\end{eqnarray}
where $\gamma$ is a dimensionless parameter representing the
probability of the quark-antiquark pair $q_3\bar{q}_4$ with
$J^{PC}=0^{++}$ creation from the vacuum, $\vec{p}_3$ and
$\vec{p}_4$ are the momenta of the created quark $q_3$ and
antiquark  $\bar{q}_4$, respectively. $\phi^{34}_{0}$,
$\omega^{34}_0$, and $\chi_{{1,-m}}^{34}$ are the flavor, color,
and spin wavefunctions of the  $q_3\bar{q}_4$, respectively. The
solid harmonic polynomial
${\cal{Y}}^m_1(\vec{p})\equiv|p|^1Y^m_1(\theta_p,\phi_p)$ reflects
the momentum-space distribution of the $q_3\bar{q}_4$ .

For the meson wavefunction, we adopt the mock meson
$|A(n_A{}^{2S_A+1}L_{A}\,\mbox{}_{J_A M_{J_A}})(\vec{P}_A)\rangle$
defined by\cite{mock}
\begin{eqnarray}
|A(n_A{}^{2S_A+1}L_{A}\,\mbox{}_{J_A M_{J_A}})(\vec{P}_A)\rangle
&\equiv& \sqrt{2E_A}\sum_{M_{L_A},M_{S_A}}\langle L_A M_{L_A} S_A
M_{S_A}|J_A
M_{J_A}\rangle\nonumber\\
&&\times  \int d^3\vec{p}_A\psi_{n_AL_AM_{L_A}}(\vec{p}_A)\chi^{12}_{S_AM_{S_A}}\phi^{12}_A\omega^{12}_A\nonumber\\
&&\times  |q_1({\scriptstyle
\frac{m_1}{m_1+m_2}}\vec{P}_A+\vec{p}_A)\bar{q}_2
({\scriptstyle\frac{m_2}{m_1+m_2}}\vec{P}_A-\vec{p}_A)\rangle,
\end{eqnarray}
where $m_1$ and $m_2$ are the masses of the quark $q_1$ with a
momentum of $\vec{p}_1$ and the antiquark $\bar{q}_2$ with a
momentum of $\vec{p}_2$, respectively. $n_A$ is the radial quantum
number of the meson $A$ composed of $q_1\bar{q}_2$.
$\vec{S}_A=\vec{s}_{q_1}+\vec{s}_{q_2}$,
$\vec{J}_A=\vec{L}_A+\vec{S}_A$, $\vec{s}_{q_1}$ ($\vec{s}_{q_2}$)
is the spin of $q_1$ ($q_2$), $\vec{L}_A$ is the relative orbital
angular momentum between $q_1$ and $q_2$.
$\vec{P}_A=\vec{p}_1+\vec{p}_2$,
$\vec{p}_A=\frac{m_1\vec{p}_1-m_1\vec{p}_2}{m_1+m_2}$. $\langle
L_A M_{L_A} S_A M_{S_A}|J_A M_{J_A}\rangle$ is a Clebsch-Gordan
coefficient, and $E_A$ is the total energy of the meson $A$.
$\chi^{12}_{S_AM_{S_A}}$, $\phi^{12}_A$, $\omega^{12}_A$, and
$\psi_{n_AL_AM_{L_A}}(\vec{p}_A)$ are the spin, flavor, color, and
space wavefunctions of the meson $A$, respectively. The mock meson
satisfies the normalization condition
\begin{eqnarray}
\langle A(n_A{}^{2S_A+1}L_{A}\,\mbox{}_{J_A M_{J_A}})(\vec{P}_A)
|A(n_A{}^{2S_A+1}L_{A}\,\mbox{}_{J_A
M_{J_A}})(\vec{P^\prime}_A)\rangle=2E_A\delta^3(\vec{P}_A-\vec{P^\prime}_A).
\end{eqnarray}
The $S$-matrix of the process $A\rightarrow BC$ is defined by
\begin{eqnarray}
\langle BC|S|A\rangle=I-2\pi i\delta(E_A-E_B-E_C)\langle
BC|T|A\rangle,
\end{eqnarray}
with
\begin{eqnarray}
\langle
BC|T|A\rangle=\delta^3(\vec{P}_A-\vec{P}_B-\vec{P}_C){\cal{M}}^{M_{J_A}M_{J_B}M_{J_C}},
\end{eqnarray}
where ${\cal{M}}^{M_{J_A}M_{J_B}M_{J_C}}$ is the helicity
amplitude of $A\rightarrow BC$. In the center of mass frame of
meson $A$, ${\cal{M}}^{M_{J_A}M_{J_B}M_{J_C}}$ can be written as
\begin{eqnarray}
{\cal{M}}^{M_{J_A}M_{J_B}M_{J_C}}(\vec{P})&=&\gamma\sqrt{8E_AE_BE_C}
\sum_{\renewcommand{\arraystretch}{.5}\begin{array}[t]{l}
\scriptstyle M_{L_A},M_{S_A},\\\scriptstyle M_{L_B},M_{S_B},\\
\scriptstyle M_{L_C},M_{S_C},m
\end{array}}\renewcommand{\arraystretch}{1}\!\!
\langle L_AM_{L_A}S_AM_{S_A}|J_AM_{J_A}\rangle\nonumber\\
&&\times\langle L_BM_{L_B}S_BM_{S_B}|J_BM_{J_B}\rangle\langle
L_CM_{L_C}S_CM_{S_C}|J_CM_{J_C}\rangle\nonumber\\
&&\times\langle 1m1-m|00\rangle\langle
\chi^{14}_{S_BM_{S_B}}\chi^{32}_{S_CM_{S_C}}|\chi^{12}_{S_AM_{S_A}}\chi^{34}_{1-m}\rangle\nonumber\\
&&\times[f_1I(\vec{P},m_1,m_2,m_3)+(-1)^{1+S_A+S_B+S_C}f_2I(-\vec{P},m_2,m_1,m_3)],
\end{eqnarray}
with $f_1=\langle
\phi^{14}_B\phi^{32}_C|\phi^{12}_A\phi^{34}_0\rangle$ and $f_2=
\langle\phi^{32}_B\phi^{14}_C|\phi^{12}_A\phi^{34}_0\rangle$,
corresponding to the contributions from Figs. 1 (a) and 1 (b),
respectively, and
\begin{eqnarray} I(\vec{P},m_1,m_2,m_3)&=&\int
d^3\vec{p}\,\mbox{}\psi^\ast_{n_BL_BM_{L_B}}
({\scriptstyle\frac{m_3}{m_1+m_2}}\vec{P}_B+\vec{p})\psi^\ast_{n_CL_CM_{L_C}}
({\scriptstyle\frac{m_3}{m_2+m_3}}\vec{P}_B+\vec{p})\nonumber\\
&&\times\psi_{n_AL_AM_{L_A}}
(\vec{P}_B+\vec{p}){\cal{Y}}^m_1(\vec{p}),
\end{eqnarray}
where $\vec{P}=\vec{P}_B=-\vec{P}_C$, $\vec{p}=\vec{p}_3$, $m_3$
is the mass of the created quark $q_3$.

The spin overlap in terms of Winger's $9j$ symbol can be given by
\begin{eqnarray}
&&\langle
\chi^{14}_{S_BM_{S_B}}\chi^{32}_{S_CM_{S_C}}|\chi^{12}_{S_AM_{S_A}}\chi^{34}_{1-m}\rangle=\nonumber\\
&&\sum_{S,M_S}\langle S_BM_{S_B}S_CM_{S_C}|SM_S\rangle\langle
S_AM_{S_A}1-m|SM_S\rangle\nonumber\\
&&(-1)^{S_C+1}\sqrt{3(2S_A+1)(2S_B+1)(2S_C+1)}\left\{\begin{array}{ccc}
\frac{1}{2}&\frac{1}{2}&S_A\\
\frac{1}{2}&\frac{1}{2}&1\\
S_B&S_C&S
\end{array}\right\}.
\end{eqnarray}

 In order to compare with experiment conventionally,
${\cal{M}}^{M_{J_A}M_{J_B}M_{J_C}}(\vec{P})$ can be converted into
the partial amplitude by a recoupling calculation\cite{recp}
\begin{eqnarray}
{\cal{M}}^{LS}(\vec{P})&=&
\sum_{\renewcommand{\arraystretch}{.5}\begin{array}[t]{l}
\scriptstyle M_{J_B},M_{J_C},\\\scriptstyle M_S,M_L
\end{array}}\renewcommand{\arraystretch}{1}\!\!
\langle LM_LSM_S|J_AM_{J_A}\rangle\langle
J_BM_{J_B}J_CM_{J_C}|SM_S\rangle\nonumber\\
&&\times\int
d\Omega\,\mbox{}Y^\ast_{LM_L}{\cal{M}}^{M_{J_A}M_{J_B}M_{J_C}}
(\vec{P}).
\end{eqnarray}
If we consider the relativistic phase space, the decay width
$\Gamma(A\rightarrow BC)$ in terms of the partial wave amplitudes
is
\begin{eqnarray}
\Gamma(A\rightarrow BC)= \frac{\pi
P}{4M^2_A}\sum_{LS}|{\cal{M}}^{LS}|^2. \label{width1}
\end{eqnarray}
Here
$P=|\vec{P}|$=$\frac{\sqrt{[M^2_A-(M_B+M_C)^2][M^2_A-(M_B-M_C)^2]}}{2M_A}$,
$M_A$, $M_B$, and $M_C$ are the masses of the meson $A$, $B$, and
$C$, respectively.

The decay width can be derived analytically if the simple harmonic
oscillator (SHO) approximation for the meson space wave functions
is used. In momentum-space, the SHO wave function is
\begin{eqnarray}
\psi_{nLM_L}(\vec{p})=R^{\mbox{\tiny
SHO}}_{nL}(p)Y_{LM_L}(\Omega_p),
\end{eqnarray}
where the radial wave function is given by
\begin{eqnarray}
R^{\mbox{\tiny SHO}}_{nL}=\frac{(-1)^n(-i)^L}{\beta^{\frac{3}{2}}}
\sqrt{\frac{2n!}{\Gamma(n+L+\frac{3}{2})}}\left(\frac{p}{\beta}\right
)^L e^{-\frac{p^2}{2\beta^2}}L^{L+\frac{1}{2}}_n({\scriptstyle
\frac{p^2}{\beta^2}}).
\end{eqnarray}
Here $\beta$ is the SHO wave function scale parameter, and
$L^{L+\frac{1}{2}}_n({\scriptstyle \frac{p^2}{\beta^2}})$ is  an
associated Laguerre polynomial.

The SHO wave functions can not be regarded as realistic, however,
they are a {\it {de facto}} standard for many nonrelativistic
quark model calculations. Moreover, the more realistic space wave
functions such as those obtained from Coulomb, plus the linear
potential model do not always result in systematic improvements
due to the inherent uncertainties of the $^3P_0$ decay model
itself\cite{3p0y,3p0x,3p0x2}. The SHO wave function approximation
is commonly employed in the $^3P_0$ decay model in literature. In
the present work, the SHO wave function approximation for the
meson space wave functions is taken.

\section{ Decays of the $3\,^1S_0$ and $4\,^1S_0$ $s\bar{s}$ states in the $^3P_0$
model}
\indent \vspace*{-1cm}

Under the SHO wave function approximation, the parameters used in
the $^3P_0$ decay model involve the $q\bar{q}$ pair production
strength parameter $\gamma$, the SHO wave function scale parameter
$\beta$, and the masses of the constituent quarks. In the present
work, we take $\gamma=6.95$ and $\beta=0.4$ GeV, the typical
values used to evaluate the light meson
decays\cite{3p0x,3p0x1,3p0x2,3p01,3p02,3p03}\footnote{Our value of
$\gamma$ is higher than that used by other groups such as
\cite{3p0x2,3p01,3p02,3p03} by a factor of $\sqrt{96\pi}$ due to
different field conventions, constant factor in $T$, etc. The
calculated results of the widths are, of course, unaffected.}, and
$m_u=m_d=0.33$ GeV, $m_s=0.55$ GeV\cite{quarkmass}. Based on the
partial wave amplitudes listed in the Appendixes A and B, and the
flavor and charge multiplicity factors shown in Table 2, from
(\ref{width1}), the numerical values of the partial decay widths
of the $4\,^1S_0$ and $3\,^1S_0$ $s\bar{s}$ states are listed in
Table 1. The initial state mass is set to 2.24 GeV and masses of
the final mesons are taken from PDG2006\cite{pdg2006} except for
the $K(2\,^3S_1)$ mass\footnote{ The assignment the $K^\ast(1410)$
as the $2\,^3S_1$ kaon is problematic\cite{3p03,vij}. Quark
model\cite{flavorfun} and other phenomenological
approaches\cite{mpla} consistently suggest the $2\,^3S_1$ kaon has
a mass about 1580 MeV, here we take 1580 MeV as the mass of the
$2\,^3S_1$ kaon.}.

\begin{table}[hbt]
\begin{center}
\caption{\small Decays of the $4\,^1S_0$ and $3\,^1S_0$ $s\bar{s}$
states in the $^3P_0$ model (In MeV). The initial state mass is
set to $2240$ MeV.} \vspace*{0.5cm}
\begin{tabular}{c|ccccccc}\hline\hline
Mode  &

$KK^\ast$ &

$K^\ast K^\ast$ &

$KK^\ast_0(1430)$&

$KK^\ast_2(1430)$&

$KK^\ast(1580)$&

$KK^\ast(1680)$ &

$\phi\phi$\\\hline

$\Gamma_i(4\,^1S_0)$&

9.1&

0.5&

1.5&

43.5&

56.9&

1.0&

12.6\\
$\Gamma_i(3\,^1S_0)$&

26.4&

8.1&

0.1&

173.3&

138.2&

0.0&

92.6
\\\hline

&\multicolumn{7}{c}{$\Gamma(4\,^1S_0)=125.1$,~~$\Gamma(3\,^1S_0)=438.7$,~~$\Gamma_{\eta(2225)}=(190\pm
30^{+40}_{-60})$}\\
         \hline\hline
\end{tabular}
\end{center}
\end{table}

 \begin{figure}[hbt]
 \begin{center}
 \epsfig{file=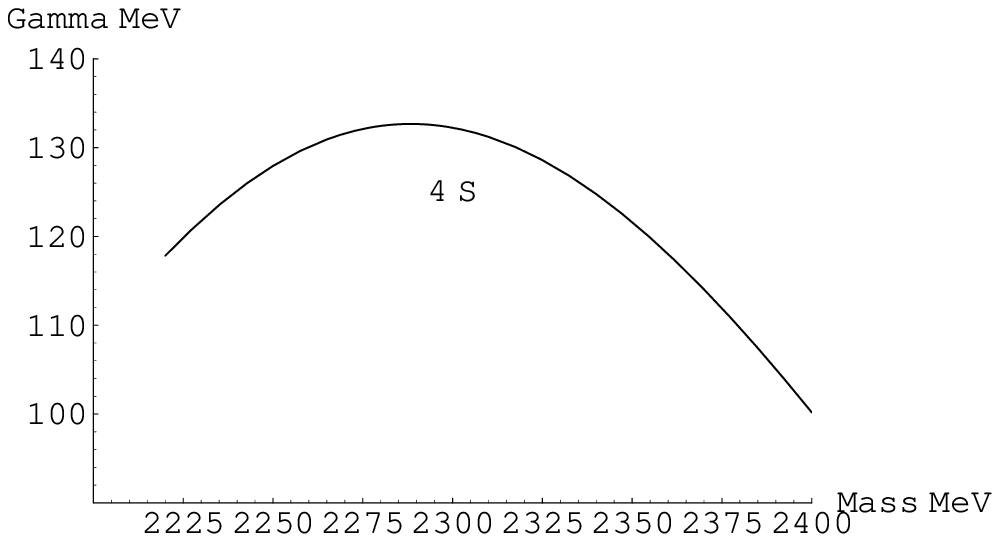,width=6.0cm, clip=}
 \epsfig{file=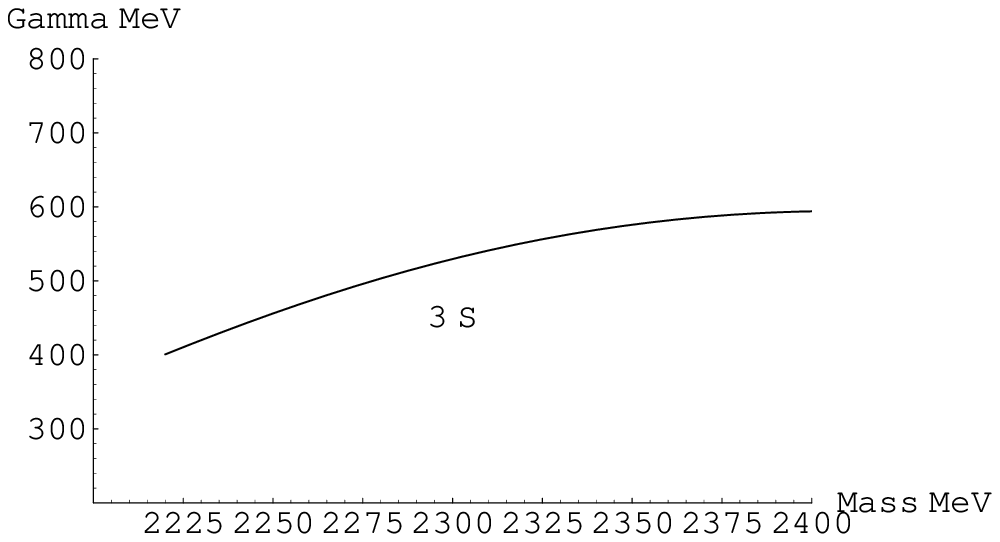,width=6.0cm, clip=}
 \caption{\small The total widths of the $4\,^1S_0$ and $3\,^1S_0$ $s\bar{s}$ states dependence on the initial state mass
 in the $^3P_0$ decay model. }
\end{center}
\end{figure}

\begin{figure}[hbt]
 \begin{center}
 \epsfig{file=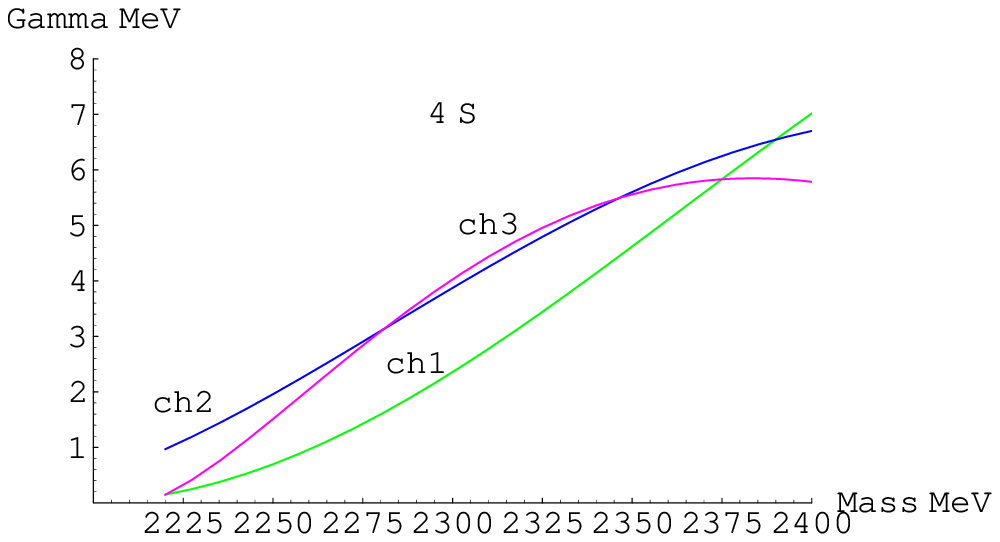, width=6.0cm, clip=}
 \epsfig{file=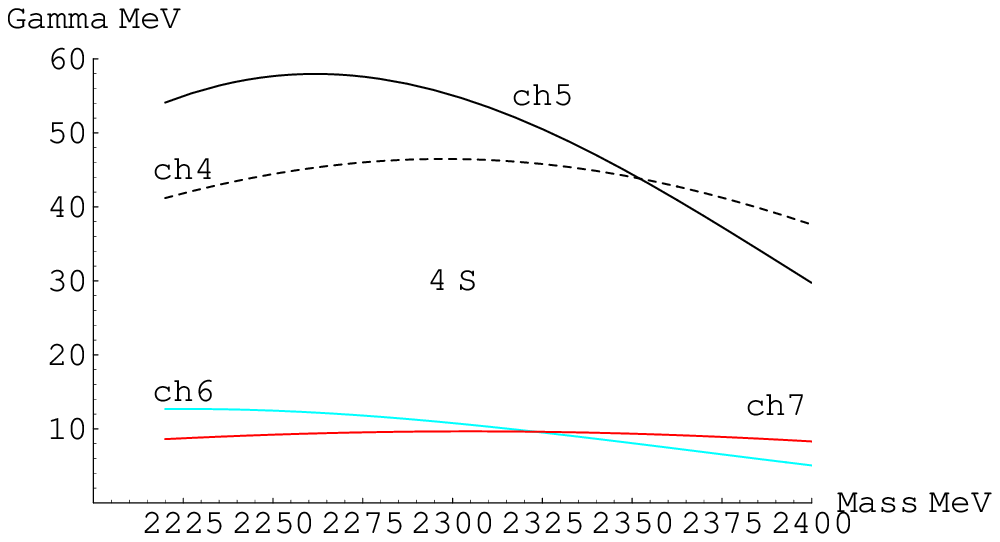, width=6.0cm, clip=}
 \epsfig{file=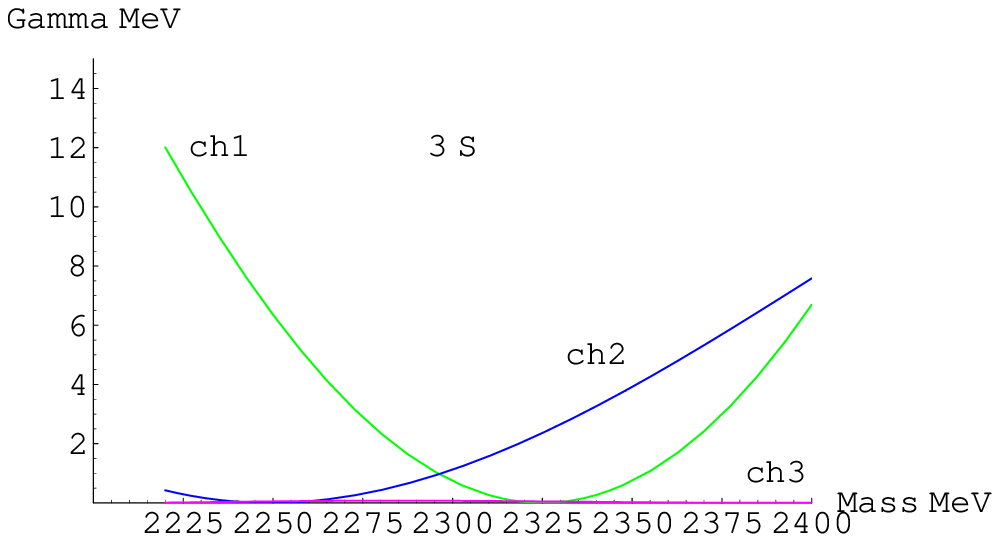, width=6.0cm, clip=}
 \epsfig{file=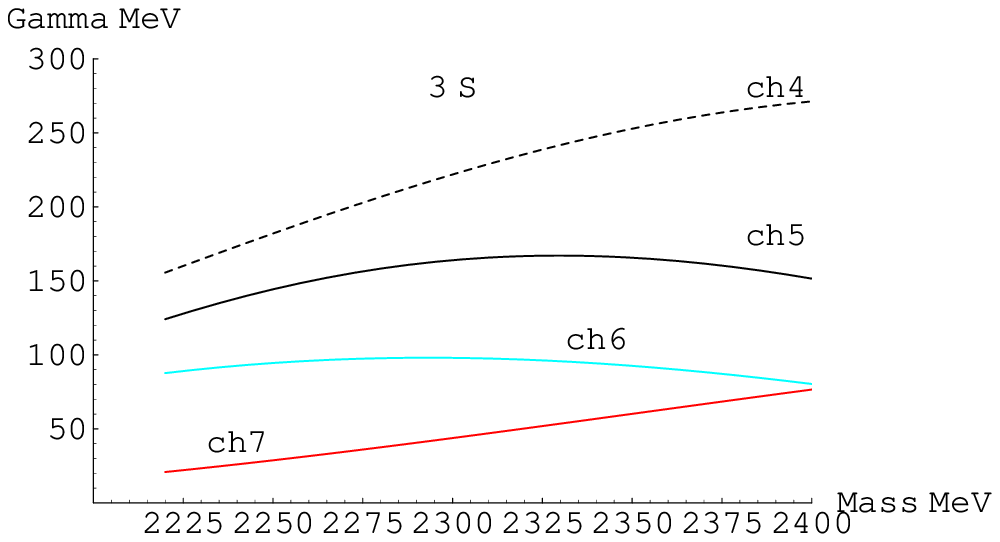, width=6.0cm, clip=}
 \caption{\small The partial widths of the $4\,^1S_0$ and $3\,^1S_0$
 $s\bar{s}$ states
 dependence on the initial state mass in the $^3P_0$ decay model. ch1=$K^\ast
 K^\ast$, ch2=$K K^\ast_0(1430)$, ch3=$K K^\ast(1680)K$, ch4=$KK^\ast_2(1430)$,
 ch5=$KK^\ast(1580)$, ch6=$\phi\phi$ and ch7=$KK^\ast$.
 }
\end{center}
\end{figure}

Table 1 indicates that the total width of the $4\,^1S_0$
$s\bar{s}$ state with a mass of 2.24 GeV predicted by the $^3P_0$
decay model is about $125.1$ MeV, consistent with the experimental
result of $\Gamma_{\eta(2225)}=(0.19\pm 0.03^{+0.04}_{-0.06})$ GeV
within errors, but the total width of the $3\,^1S_0$ $s\bar{s}$
state with a mass of 2.24 GeV is predicted to be about $438.7$
MeV, incompatible with the measured width of the $\eta(2225)$.
Also, in order to check the dependence of the predicted results on
the initial state mass, the variation of the total widths of the
$4\,^1S_0$ and $3\,^1S_0$ $s\bar{s}$ states with the initial state
mass is shown in Fig. 2.  From Fig. 2, we can see that when the
initial state mass varies from 2220 to 2400 MeV, the total width
of the $4\,^1S_0$ $s\bar{s}$ state varies from about 100 to 132
MeV, lying in the width range of the $\eta(2225)$, while the total
width of the $3\,^1S_0$ $s\bar{s}$ state varies from about 400 to
594 MeV, far more than the width of the $\eta(2225)$. Therefore,
it would be very difficult to identify the $\eta(2225)$ as the
$3\,^1S_0$ $s\bar{s}$ state, but the assignment of the
$\eta(2225)$ as the $4\,^1S_0$ $s\bar{s}$ state seams reasonable
for accounting for the total width of the $\eta(2225)$, assuming
the $^3P_0$ meson decay model is accurate.

The variation of the partial decay widths of the $4\,^1S_0$ and
$3\,^1S_0$ $s\bar{s}$ states with the initial state mass is also
shown in Fig. 3. For both $4\,^1S_0$ and $3\,^1S_0$ $s\bar{s}$
states, the partial widths of the modes $\phi\phi$ and $KK^\ast$
depend weakly on the initial state mass, while the partial widths
of the modes $K^\ast K^\ast$ and $K K^\ast_0(1430)$ vary
dramatically with the initial state mass, and the
$KK^\ast_2(1430)$ and $KK^\ast(1580)$ modes always have a sizable
branch ratio in the mass region of the $\eta(2225)$. It is
interesting to note that, in the mass region $2.22\sim 2.40$ GeV,
$\Gamma(3\,^1S_0\rightarrow KK^\ast(1680))$ is almost 0 MeV, while
$\Gamma(4\,^1S_0\rightarrow KK^\ast(1680))$ varies from 0.15 MeV
to 5.8 MeV.

\section{Summary and conclusion}
\indent \vspace*{-1cm}

The strong decays of the $3\,^1S_0$ and $4\,^1S_0$ $s\bar{s}$
states in the $^3P_0$ meson decay model indicates that if the
initial state mass is set to 2.24 GeV, the central value of the
$\eta(2225)$ mass measured by the BES Collaboration\cite{BES08},
the total widths of the $3\,^1S_0$ and $4\,^1S_0$ $s\bar{s}$
states are predicted to be about 438 MeV and 125 MeV,
respectively. Also, the variation of the total widths of the
$4\,^1S_0$  and $3\,^1S_0$ $s\bar{s}$ states with the initial
state mass shows that, in the mass region of the $\eta(2225)$, the
total width of the $4\,^1S_0$ $s\bar{s}$ state lies in the range
about $100\sim 132 $ MeV, while the total width of the $3\,^1S_0$
$s\bar{s}$ state lies in the range about $400\sim 594$ MeV. A
comparison of the $^3P_0$ model predictions and
$\Gamma_{\eta(2225)}=(0.19\pm 0.03^{+0.04}_{-0.06})$ GeV reported
by the BES Collaboration\cite{BES08} indicates that it would be
very difficult to identify the $\eta(2225)$ as the $3\,^1S_0$
$s\bar{s}$ state, while the assignment of the $\eta(2225)$ as the
$4\,^1S_0$ $s\bar{s}$ state seems reasonable to reproduce the
total width of the $\eta(2225)$. We therefore tend to conclude
that the $\eta(2225)$ may be a good candidate for the $4\,^1S_0$
$s\bar{s}$ state.

 \section*{Acknowledgments}
 This work
is supported in part by HANCET under Contract No. 2006HANCET-02,
and Program for Youthful Teachers in University of Henan Province.

\appendix
\newcounter{zaehler}
\renewcommand{\thesection}{\Alph{zaehler}}
\renewcommand{\theequation}{\Alph{zaehler}.\arabic{equation}}
\setcounter{equation}{0} \addtocounter{zaehler}{1}
\section*{Appendix A: The amplitudes for the $4\,^1S_0$ $q\bar{q}$ decay in ${^3P_0}$
model} \vspace*{-0.8cm} {\tiny
\begin{eqnarray}
&&{\cal{M}}^{LS}(4^1S_0\rightarrow 1^3S_1+1^3S_1)=\nonumber\\
&& \gamma
e^{-\frac{[m_1m_2(m_2-m_3)m_3+m^2_2m^2_3+m^2_1(m^2_2+m_2m_3+m^2_3)]P^2}{3\beta^2(m_1+m_3)^2(m_2+m_3)^2}}
 \sqrt{E_aE_bE_c}\frac{1}{\pi^{3/4}}(f_1+f_2) P\nonumber\\
&&\times \left
[8505\beta^6(m_1+m_3)^6(m_2+m_3)^6(3m_1m_2+2m_1m_3+2m_2m_3+m^2_3)
-1134\beta^4(m_1+m_3)^4\right.\nonumber\\
&&~~~~\times
(m_2+m_3)^4(7m_1m_2+6m_1m_3+6m_2m_3+5m^2_3)(m_2m_3+2m_1m_2+m_1m_3)^2P^2\nonumber\\
&&~~~~\left.+108\beta^2(m_1+m_3)^2(m_2+m_3)^2(5m_1m_2+6m_1m_3+6m_2m_3+7m^2_3)(m_2m_3+2m_1m_2+m_1m_3)^4P^4\right.\nonumber\\
&&~~~~\left.-8(m_2m_3+2m_1m_2+m_1m_3)^6(m_1m_2+2m_1m_3+2m_2m_3+3m^2_3)P^6 \right ]\nonumber\\
 &&\times\frac{\sqrt{2}}{19683\sqrt{105}\beta^{15/2}}\frac{1}{(m_1+m_3)^7(m_2+m_3)^7}\\
&&{\cal{M}}^{LS}(4^1S_0\rightarrow 1^3S_1+1^1S_0)=\nonumber\\
&& \gamma
e^{-\frac{[m_1m_2(m_2-m_3)m_3+m^2_2m^2_3+m^2_1(m^2_2+m_2m_3+m^2_3)]P^2}{3\beta^2(m_1+m_3)^2(m_2+m_3)^2}}
 \sqrt{E_aE_bE_c}\frac{1}{\pi^{3/4}}(f_1-f_2) P\nonumber\\
&&\times \left
[8505\beta^6(m_1+m_3)^6(m_2+m_3)^6(3m_1m_2+2m_1m_3+2m_2m_3+m^2_3)
-1134\beta^4(m_1+m_3)^4\right.\nonumber\\
&&~~~~\times
(m_2+m_3)^4(7m_1m_2+6m_1m_3+6m_2m_3+5m^2_3)(m_2m_3+2m_1m_2+m_1m_3)^2P^2\nonumber\\
&&~~~~\left.+108\beta^2(m_1+m_3)^2(m_2+m_3)^2(5m_1m_2+6m_1m_3+6m_2m_3+7m^2_3)(m_2m_3+2m_1m_2+m_1m_3)^4P^4\right.\nonumber\\
&&~~~~\left.-8(m_2m_3+2m_1m_2+m_1m_3)^6(m_1m_2+2m_1m_3+2m_2m_3+3m^2_3)P^6 \right ]\nonumber\\
 &&\times\frac{1}{19683\sqrt{105}\beta^{15/2}}\frac{1}{(m_1+m_3)^7(m_2+m_3)^7}\\
&&{ \cal{M}}^{LS}(4^1S_0\rightarrow 1^3P_0+1^1S_0)=i\gamma
e^{\frac{-[(m_1m_2(m_2-m_3)m_3+m^2_2m^2_3+m^2_1(m^2_2+m_2m_3+m^2_3)]P^2}{3\beta^2(m_1+m_3)^2(m_2+m_3)^2}}\sqrt{E_aE_bE_c}\frac{1}{\pi^{3/4}}\nonumber\\
&& \times\left\{(f_1+f_2)\frac{\sqrt{70}}{81\beta^{1/2}}\right.\nonumber\\
&&~~+\left.\left [(10m^2_2m^2_3+6m_2m^3_3+19m^2_1m^2_2+19m^2_1m_2m_3+4m^2_1m^2_3+28m_1m^2_2m_3+23m_1m_2m^2_3+3m_1m^3_3)f_2\right.\right.\nonumber\\
&&~~~~+\left.\left.(4m^2_2m^2_3+3m_2m^3_3+19m_1m^2_2m_3+23m_1m_2m^2_3+6m_1m^3_3+19m^2_1m^2_2+28m^2_1m_2m_3+10m^2_1m^2_3)f_1\right]\right.\nonumber\\
&&~~~~\times\frac{-\sqrt{70}}{729\beta^{5/2}}\frac{1}{(m_1+m_3)^2(m_2+m_3)^2}P^2\nonumber\\
&&~~+\left [(m_2+m_3)(5m^2_1m_2+6m_2m^2_3+12m_1m_2m_3+m_1m^2_3)f_2+(m_1+m_3)(m_2m^2_3+5m_1m^2_2+12m_1m_2m_3+6m_1m^2_3)f_1\right]\nonumber\\
&&~~~~\times(m_2m_3+2m_1m_2+m_1m_3)^2\frac{2\sqrt{14}}{729\sqrt{5}\beta^{9/2}}\frac{1}{(m_1+m_3)^4(m_2+m_3)^4}P^4\nonumber\\
&&~~+\left [(14m^2_2m^2_3+18m_2m^3_3+5m^2_1m^2_2+5m^2_1m_2m_3-4m^2_1m^2_3+20m_1m^2_2m_3+25m_1m_2m^2_3-3m_1m^3_3)f_2\right.\nonumber\\
&&~~~~+\left.(-4m^2_2m^2_3-3m_2m^3_3+5m^2_1m^2_2+20m^2_1m_2m_3+14m^2_1m^2_3+5m_1m^2_2m_3+25m_1m_2m^2_3+18m_1m^3_3)f_1\right]\nonumber\\
&&~~~~\left.\times(m_2m_3+2m_1m_2+m_1m_3)^4\frac{-4\sqrt{2}}{6561\sqrt{35}\beta^{13/2}}\frac{1}{(m_1+m_3)^6(m_2+m_3)^6}P^6\right.\nonumber\\
&&~~+\left [
(m_1m_2-m_1m_3+2m_2m_3)f_2+(m_1m_2+2m_1m_3-m_2m_3)f_1\right
](m_2m_3+2m_1m_2+m_1m_3)^6\nonumber\\
&&~~~~\times\left.(m_1m_2+2m_1m_3+2m_2m_3+3m^2_3)\frac{8\sqrt{2}}{177147\sqrt{35}\beta^{17/2}}\frac{1}{(m_1+m_3)^8(m_2+m_3)^8}P^8\right\}\\
&&{ \cal{M}}^{LS}(4^1S_0\rightarrow 1^3P_2+1^1S_0)=i\gamma
e^{\frac{-[(m_1m_2(m_2-m_3)m_3+m^2_2m^2_3+m^2_1(m^2_2+m_2m_3+m^2_3)]P^2}{3\beta^2(m_1+m_3)^2(m_2+m_3)^2}}\sqrt{E_aE_bE_c}\frac{1}{\pi^{3/4}}\nonumber\\
&&\times\left\{
4032(f_1+f_2)\beta^{18}\left(\frac{m_1}{m_1+m_3}+\frac{m_2}{m_2+m_3}\right)^2P^2\frac{1}{19683\sqrt{35}\beta^{41/2}}\right.\nonumber\\
&&~~+\left
[\left(2m_2m^2_3(10m_2+9m_3)+m^2_1(17m^2_2+17m_2m_3+2m^2_3)+m_1m_3(44m^2_2+49m_2m_3+9m^2_3)\right)35\beta^2f_2\right.\nonumber\\
&&~~~~\left.\left(
m_2m^2_3(2m_2+9m_3)+m_1m_3(17m^2_2+49m_2m_3+18m^2_3)+m^2_1(17m^2_2+44m_2m_3+20m^2_3)\right)35\beta^2f_1\right.\nonumber\\
&&~~~~-\left.128(f_1+f_2)m^2_1m^2_2P^2\right]\frac{2}{2187\sqrt{35}\beta^{9/2}}\frac{1}{(m_1+m_3)^2(m_2+m_3)^2}P^2\nonumber\\
&&~~+\left [\left (
2m_2m^2_3(26m_2+27m_3)+m^2_1(37m^2_2+37m_2m_3-2m^2_3)+m_1m_3(100m^2_2+113m_2m_3+9m^2_3)\right )f_2\right.\nonumber\\
&&~~~~\left.\left (
m_2m^2_3(9m_3-2m_2)+m^2_1(37m^2_2+100m_2m_3+52m^2_3)+m_1m_3(37m^2_2+113m_2m_3+54m^2_3)\right )f_1\right ]\nonumber\\
&&~~~~\times(m_2m_3+2m_1m_2+m_1m_3)^2\frac{-4\sqrt{7}}{6561\sqrt{5}\beta^{9/2}}\frac{1}{(m_1+m_3)^4(m_2+m_3)^4}P^4\nonumber\\
&&~~ + \left [\left (
(2m_2m^2_3(22m_2+27m_3)+m^2_1(23m^2_2+23m_2m_3-10m^3_2)+m_1m_3(68m^2_2+79m_2m_3-9m^2_3)\right)f_2\right.\nonumber\\
&&~~~~\left.\left ((m^2_1(23m^2_2+68m_2m_3+44m^2_3)-m_2m^2_3(10m_2+9m_3)+m_1m_3(23m^2_2+79m_2m_3+54m^2_3)\right)f_1\right ]\nonumber\\
&&~~~~\times(m_2m_3+2m_1m_2+m_1m_3)^4\frac{8}{19683\sqrt{35}\beta^{13/2}}\frac{1}{(m_1+m_3)^6(m_2+m_3)^6}P^6\nonumber\\
&&~~+\left[
(m_1m_2-m_1m_3+2m_2m_3)f_2+(m_1m_2+2m_1m_3-m_2m_3)f_1\right](m_2m_3+2m_1m_2+m_1m_3)^6\nonumber\\
&&~~~~\times\left.(m_1m_2+2m_1m_3+2m_2m_3+3m^2_3)
\frac{-16}{177147\sqrt{35}\beta^{17/2}}\frac{1}{(m_1+m_3)^8(m_2+m_3)^8}P^8\right\}\\
&&{ \cal{M}}^{LS}(4^1S_0\rightarrow 2^3S_1+1^1S_0)=\gamma
e^{\frac{-[(m_1m_2(m_2-m_3)m_3+m^2_2m^2_3+m^2_1(m^2_2+m_2m_3+m^2_3)]P^2}{3\beta^2(m_1+m_3)^2(m_2+m_3)^2}}\sqrt{E_aE_bE_c}\frac{1}{\pi^{3/4}}P(f_2-f_1)\nonumber\\
&&\times\left\{\left[25515\beta^8(m_1+m_3)^8(m_2+m_3)^8(53m_1m_2+46m_1m_3+34m_2m_3+27m^2_3)\right.\right.\nonumber\\
&&~~~~-3402\beta^6(m_1+m_3)^6(m_2+m_3)^6\left
(20m^2_2m^3_2(2m_2+3m_3)+61m_1m_2m^2_3(46m^2_2+91m_2m_3+35m^2_3)\right)\nonumber\\
&&~~~~+\left.
3m^2_1m_3(208m^3_2+497m^2_2m_3+334m_2m^2_3+65m^3_3)+m^3_1(419m^3_2+1098m^2_2m_3+885m_2m^2_3+226m^3_3)\right]P^2\nonumber\\
&&~~+\left [
2m_2m^2_3(4m_2+21m_3)+m^2_1(95m^2_2+263m_2m_3+134m^2_3)+m_1m_3(116m^2_2+331m_2m_3+147m^2_3)\right
] \nonumber\\
&&~~~~\times
927\beta^4m_1(m_1+m_3)^4(m_2+m_3)^4(m_2m_3+2m_1m_2+m_1m_3)^2P^4\nonumber\\
&&~~-\left[
16m^3_2m^3_3-6m_1m_2m^2_3(16m^2_2+51m_2m_3+27m^2_3)+3m^2_1m_3(24m^3_2+139m^2_2m_3+266m_2m^2_3+135m^3_3)\right.\nonumber\\
&&~~~~+\left.m^3_1(89m^3_2+438m^2_2m_3+675m_2m^2_3+310m^3_3)\right
]24\beta^2(m_1+m_3)^2(m_2+m_3)^3(m_2m_3+2m_1m_2+m_1m_3)^4P^6\nonumber\\
&&~~+\left.16(m_2m_3+2m_1m_2+m_1m_3)^6(m_2m_3-m_1m_2-2m_1m_3)^2(m_1m_2+2m_1m_3+2m_2m_3+3m^2_3)P^8\right\}\nonumber\\
&&\times\frac{1}{531441\sqrt{70}\beta^{19/2}}\frac{1}{(m_1+m_3)^9(m_2+m_3)^9}\\
&&{ \cal{M}}^{LS}(4^1S_0\rightarrow 1^3D_1+1^1S_0)=\gamma
e^{\frac{-[(m_1m_2(m_2-m_3)m_3+m^2_2m^2_3+m^2_1(m^2_2+m_2m_3+m^2_3)]P^2}{3\beta^2(m_1+m_3)^2(m_2+m_3)^2}}\sqrt{E_aE_bE_c}\frac{1}{\pi^{3/4}}\nonumber\\
&&\times\left\{\left [\left
(119070-\frac{11907m_2(m_2+4m_3)P^2}{\beta^2(m_2+m_3)^2}-\frac{162m^3_2(43m_2+18m_3)P^4}{\beta^4(m_2+m_3)^4}+\frac{4m^5_2(89m_2+216m_3)P^6}{\beta^6(m_2+m_3)^6}+\frac{8m^7_2(m_2+2m_3)P^8}{\beta^8(m_2+m_3)^8}\right)f_1\right.\right.\nonumber\\
&&~~~~\left.\left(8505+\frac{3402m_2(15m_2+14m_3)P^2}{\beta^2(m_2+m_3)^2}-\frac{324m^3_2(97m_2+138m_3)P^4}{\beta^4(m_2+m_3)^4}+\frac{8m^5_2(443m_2+702m_3)P^6}{\beta^6(m_2+m_3)^6}-\frac{32m^7_2(3m_2+5m_3)P^8}{\beta^8(m_2+m_3)^8}\right)f_2\right ]\nonumber\\
&&~~~~\times\frac{-2\sqrt{2}}{885735\sqrt{7}\beta^{3/2}}\frac{m_1P}{m_1+m_3}\nonumber\\
&&~~+\left[2\left(-1701\beta^6(m_2+m_3)^6(23m_2+22m_3)+81\beta^4m^2_2(m_2+m_3)^4(32m_2+141m_3)P^2\right.\right.\nonumber\\
&&~~~~~+\left.12\beta^2m^4_2(m_2+m_3)^2(37m_2+63m_3)P^4-4m^6_2(4m_2+5m_3)P^6\right)f_1\nonumber\\
&&~~+\left(-1701\beta^6(m_2+m_3)^6(22m_2+m_3)+486\beta^4m^2_2(2m_2-47m_3)(m_2+m_3)^4P^2\right.\nonumber\\
&&~~~~~+\left.\left.12\beta^2m^4_2(m_2+m_3)^2(214m_2+549m_3)P^4-8m^6_2(18m_2+37m_3)P^6\right)f_2\right]\frac{-2\sqrt{2}}{885735\sqrt{7}\beta^{19/2}}\frac{m^2_1P^3}{(m_2+m_3)^7(m_1+m_3)^2}\nonumber\\
&&~~+\left[-567\beta^6(40f_1+29f_2)+\frac{162\beta^4m_2(129m_2f_1+276m_3f_1+127m_2f_2+18m_3f_2)P^2}{(m_2+m_3)^2}\right.\nonumber\\
&&~~~~-\left.\frac{4\beta^2m^3_2(218m_2f_1+468m_3f_1+427m_2f_2-468m_3f_2)P^4}{(m_2+m_3)^4}-\frac{8m^5_2(9m_2f_1+16m_3f_1+5m_2f_2+26m_3f_2)P^6}{(m_2+m_3)^6}\right]\nonumber\\
&&~~~~\times\frac{-2\sqrt{2}}{885735\sqrt{7}\beta^{19/2}}\frac{m^3_1P^3}{(m_1+m_3)^3}\nonumber\\
&&~~~~+\left[81\beta^4(m_2+m_3)^4(50m_2f_1+132m_3f_1+40m_2f_2+15m_3f_2)-2\beta^2m^2_2(m_2+m_3)^2(752m_2f_1+1647m_3f_1+628m_2f_2+378m_3f_2)P^2\right.\nonumber\\
&&~~~~+\left.
4m^4_2(2m_2f_1-5m_3f_1+12m_2f_2+5m_3f_2)P^4\right]\frac{-4\sqrt{2}}{885735\sqrt{7}\beta^{19/2}}\frac{m^4_1P^5}{(m_2+m_3)^5(m_1+m_3)^4}\nonumber\\
&&~~+\left[81\beta^4(16f_1+3f_2)(m_2+m_3)^4+6\beta^2m_2(m_2+m_3)^2(133m_2f_1+468m_3f_1+20m_2f_2+72m_3f_2)P^2\right.\nonumber\\
&&~~~-\left. 4m^3_2(19m_2f_1+26m_3f_1+9m_2f_2+16m_3f_2)P^4\right
]\frac{4\sqrt{2}}{885735\sqrt{7}\beta^{19/2}}\frac{m^5_1P^5}{(m_2+m_3)^4(m_1+m_3)^5}\nonumber\\
&&~~+\left
[\beta^2(m_2+m_3)^2(122m_2f_1-396m_3f_1+118m_2f_2-9m_3f_2)+2m^2_2(16m_2f_1+37m_3f_1-2m_2f_2+5m_3f_2)P^2\right.\nonumber\\
&&~~~~\times\frac{-8\sqrt{2}}{885735\sqrt{7}\beta^{19/2}}\frac{m^6_1P^7}{(m_2+m_3)^3(m_1+m_3)^6)}\nonumber\\
&&~~+\left[
\beta^2(104f_1+25f_2)(m_2+m_3)^2+2m_2(m_2f_1+20m_3f_1-3m_2f_2-2m_3f_2)P^2\right
]\frac{-8\sqrt{2}}{885735\sqrt{7}\beta^{19/2}}\frac{m^7_1P^7}{(m_2+m_3)^2(m_1+m_3)^7}\nonumber\\
&&~~+\left[4(m_2-m_3)f_1+m_3f_2\right]\frac{16\sqrt{2}}{885735\sqrt{7}\beta{19/2}}\frac{m^8_1P^9}{(m_2+m_3)(m_1+m_3)^8}+(4f_1-f_2)\frac{16\sqrt{2}}{2657205\sqrt{7}\beta^{19/2}}\frac{m^9_1P^9}{(m_1+m_3)^9}\nonumber\\
&&~~+\left[\left(
-25515(m_2+m_3)^8+\frac{49329m^2_2(m_2+m_3)^6P^2+5103m_2(m_2+m_3)^7P^2}{\beta^2}\right.\right.\nonumber\\
&&~~~~+\left.\frac{1458m^4_2(m_2+m_3)^4P^4-7290m^3_2(m_2+m_3)^5P^4}{\beta^4}+\frac{108m^5_2(m_2+m_3)^3P^6-300m^6_2(m_2+m_3)^2P^6}{\beta^6}\right.\nonumber\\
&&~~~~+\left.\frac{24m^7_2(m_2+m_3)P^8-8m^8_2P^8}{\beta^8}\right)m_2f_1\nonumber\\
&&~~+ \left(
-178605(m_2+m_3)^8+\frac{34020m^2_2(m_2+m_3)^6P^2+112266m_2(m_2+m_3)^7P^2}{\beta^2}\right.\nonumber\\
&&~~~~+\left.\frac{3888m^4_2(m_2+m_3)^4P^4-32076m^3_2(m_2+m_3)^5P^4}{\beta^4}+\frac{2376m^5_2(m_2+m_3)^3P^6-624m^6_2(m_2+m_3)^2P^6}{\beta^6}\right.\nonumber\\
&&~~~~+\left.\left.\left.\frac{16m^8_2P^8-48m^7_2(m_2+m_3)P^8}{\beta^8}\right)2m_2f_2\right]\frac{-2\sqrt{2}}{2657205\sqrt{7}\beta^{3/2}}\frac{P}{(m_2+m_3)^9}\right\}
\end{eqnarray} }
\setcounter{equation}{0} \addtocounter{zaehler}{1}
\section*{Appendix B: The amplitudes for the $3\,^1S_0$ $q\bar{q}$ decay in $^3P_0$
model} \vspace*{-0.8cm} {\tiny
\begin{eqnarray}
&&{ \cal{M}}^{LS}(3^1S_0\rightarrow 2^3S_1+1^1S_0)=\gamma
e^{\frac{-[(m_1m_2(m_2-m_3)m_3+m^2_2m^2_3+m^2_1(m^2_2+m_2m_3+m^2_3)]P^2}{3\beta^2(m_1+m_3)^2(m_2+m_3)^2}}\sqrt{E_aE_bE_c}\frac{1}{\pi^{3/4}}(f_1-f_2)P\nonumber\\
&&\left \{ \left [
405\beta^6(m_1+m_3)^6(m_2+m_3)^6(63m_1m_2+70m_1m_3+50m_2m_3+57m^2_3)-54\beta^4(m_1+m_3)^4(m_2+m_3)^4\right.\right.\nonumber\\
&&~~~~\times\left (22m^3_2m^3_2+45m^2_2m^4_3+m_1m_2m^2_3(163m^2_2+472m_2m_3+240m^2_3)+m^3_1(285m^3_2+1022m^2_2m_3+1022m_2m^2_3+308m^3_3)\right.\nonumber\\
&&~~~~\left.\left.+m^2_1m_3(448m^3_2+1515m^2_2m_3+1328m_2m^2_3+330m^3_3)\right)\right ]P^2\nonumber\\
&&~~+\left [
36\beta^2(m_1+m_3)^2(m_2+m_3)^2(m_2m_3+2m_1m_2+m_1m_3)^2\left
(m^2_2(2m_2-3m_3)m^3_3-m_1m_2m^2_3(13m^2_2+46m_2m_3+18m^2_3)\right.\right.\nonumber\\
&&~~~~\left.\left.+
m^3_1(18m^3_2+88m^2_2m_3+133m_2m^2_3+58m^3_3)+m^2_1m_3(20m^3_2+96m^2_2m_3+166m_2m^2_3+75m^3_3)\right)\right
]P^4\nonumber\\
&&~~\left.-8(m_2m_3+2m_1m_2+m_1m_3)^4(m_2m_3-m_1m_2-2m_1m_3)^2(m_1m_2+2m_1m_3+2m_2m_3+3m^3_3)P^6\right
\}\nonumber\\
&&~~\times\frac{1}{19683\sqrt{15}\beta^{15/2}}\frac{1}{(m_1+m_3)^7(m_2+m_3)^7}\\
&&{ \cal{M}}^{LS}(3^1S_0\rightarrow 1^3D_1+1^1S_0)=4\gamma
e^{\frac{-[(m_1m_2(m_2-m_3)m_3+m^2_2m^2_3+m^2_1(m^2_2+m_2m_3+m^2_3)]P^2}{3\beta^2(m_1+m_3)^2(m_2+m_3)^2}}\sqrt{E_aE_bE_c}\frac{1}{\pi^{3/4}}P\nonumber\\
&&\left \{ \left [ \left(
405\beta^6(m_1+m_3)^6(m_2+m_3)^6(3m_1m_2+14m_1m_3-11m_2m_3)+27\beta^4(m_1+m_3)^4(m_2+m_3)^4(50m^3_2m^3_3+33m^2_2m^4_3)\right.\right.\right.\nonumber\\
&&~~~+m_1m_2m^2_3(203m^2_2+68m_2m_3-84m^2_3)+m^3_1(27m^3_2-374m^2_2m_3-608m_2m^2_3-224m^3_3)\nonumber\\
&&~~~~\left.-m^2_1m_3(-152m^3_2+423m^2_2m_3+776m_2m^2_3+252m^3_3)\right
)P^2+
36\beta^2(m_1+m_3)^2(m_2+m_3)^2(m_2m_3+2m_1m_2+m_1m_3)^2\nonumber\\
&&~~~~\times\left (
m^2_2(m_2-3m_3)m^3_3-4m_1m_2m^2_3(2m^2_2+8m_2m_3+3m^2_3)+m^3_1(3m^3_2+26m^2_2m_3+53m_2m^2_3+26m^3_3)\right.\nonumber\\
&&~~~~+\left. m^2_1m_3(4m^3_2+27m^2_2m_3+71m_2m^2_3+36m^3_3)\right
) P^4-\left(
4(m_2m_3+2m_1m_2+m_1m_3)^4(m_2m_3-m_1m_2+2m_1m_3)^2\right.\nonumber\\
&&~~~~\left.\left.\times(m_1m_2+2m_1m_3+2m_2m_3+3m^2_3)\right)P^6\right
]f_1\nonumber\\
&&+\left [ \left(
405\beta^6(m_1+m_3)^6(m_2+m_3)^6(-3m_1m_2-14m_2m_3+11m_1m_3)+27\beta^4(m_1+m_3)^4(m_2+m_3)^4(224m^3_2m^3_3+252m^2_2m^4_3)\right.\right.\nonumber\\
&&~~~+4m_1m_2m^3_2(152m^2_2+194m_2m_3+21m^2_3)-m^3_1(27m^3_2+152m^2_2m_3+203m_2m^2_3+50m^3_3)\nonumber\\
&&~~~~\left.+m^2_1m_3(374m^3_2+423m^2_2m_3-68m_2m^2_3-33m^3_3)\right
)P^2-
36\beta^2(m_1+m_3)^2(m_2+m_3)^2(m_2m_3+2m_1m_2+m_1m_3)^2\nonumber\\
&&~~~~\times\left
(2m^2_2m^3_3(13m_2+18m_3)+m_1m_2m^2_3(53m^2_2+71m_2m_3-12m^2_3)+m^3_1(3m^3_2+4m^2_2m_3-8m_2m^2_3+m^3_3)\right.\nonumber\\
&&~~~~\left.+ m^2_1m_3(26m^3_2+27m^2_2m_3-32m_2m^2_3-3m^3_3)\right
) P^4+\left(
4(m_1m_2-m_1m_3+2m_2m_3)^2(m_2m_3+2m_1m_2+m_1m_3)^4\right.\nonumber\\
&&~~~~\left.\left.\left.\times(m_1m_2+2m_1m_3+2m_2m_3+3m^2_3)\right)P^6\right
]f_2
\right\}\frac{1}{98415\sqrt{3}\beta^{15/2}}\frac{1}{(m_1+m_3)^7(m_2+m_3)^7}
\end{eqnarray}
} The amplitudes of  $3^1S_0\rightarrow 1^3S_1+1^3S_1$,
$1^3S_1+1^1S_0$, $1^3P_0+1^1S_0$ and $1^3P_2+1^1S_0$ are taken
from Appendix A of Ref.\cite{eta1835}.

 \vspace*{-1cm}
\section*{Appendix C: Flavor and charge multiplicity factors}
\indent \vspace*{-0.5cm}

The flavor factors $f_1$ and $f_2$  can be calculated using the
matrix notation introduced in Ref.\cite{3p0rev3} with the meson
flavor wavefunctions following the conventions of
Ref.\cite{flavorfun} for the special process with definite charges
like $s\bar{s}\rightarrow K^{\ast+}K^-$. In order to obtain the
general (i.e. charge independent) width of decays like
$s\bar{s}\rightarrow K^\ast K$, one should multiply the width
$\Gamma(s\bar{s}\rightarrow K^{\ast +}K^-)$ by a charge
multiplicity factor ${\cal{F}}$. The $f_1$, $f_2$ and ${\cal{F}}$
for all the processes considered in this work are given in Table
2.
\begin{table}[hbt]
\begin{center}
\caption{\small Flavor and charge multiplicity factors}
\vspace*{0.5cm}
\begin{tabular}{ccccc}\hline\hline
General decay & subprocess& $f_1$& $f_2$ & ${\cal{F}}$\\\hline

$s\bar{s}\rightarrow K^\ast K$ &

$s\bar{s}\rightarrow K^{\ast +}K^-$ &

$0$ &

$-\frac{1}{\sqrt{3}}$&

$4$\\

$s\bar{s}\rightarrow K^\ast_0(1430) K$ &

$s\bar{s}\rightarrow K^{\ast +}_0(1430)K^-$ &

$0$ &

$-\frac{1}{\sqrt{3}}$&

$4$\\

$s\bar{s}\rightarrow K^\ast_2(1430) K$ &

$s\bar{s}\rightarrow K^{\ast +}_2(1430)K^-$ &

$0$ &

$-\frac{1}{\sqrt{3}}$&

$4$\\

$s\bar{s}\rightarrow K^\ast(1580) K$ &

$s\bar{s}\rightarrow K^{\ast +}(1580)K^-$ &

$0$ &

$-\frac{1}{\sqrt{3}}$&

$4$\\

$s\bar{s}\rightarrow K^\ast(1680) K$ &

$s\bar{s}\rightarrow K^{\ast +}(1680)K^-$ &

$0$ &

$-\frac{1}{\sqrt{3}}$&

$4$\\

$s\bar{s}\rightarrow K^\ast K^\ast$

&$s\bar{s}\rightarrow K^{\ast+}K^{\ast -}$ &

$0$ &

$-\frac{1}{\sqrt{3}}$ &

$2$\\

$s\bar{s}\rightarrow \phi\phi $ &

$s\bar{s}\rightarrow \phi\phi$ &

$+\frac{1}{\sqrt{3}}$ &

$+\frac{1}{\sqrt{3}}$&

$\frac{1}{2}$
\\\hline\hline
\end{tabular}
\end{center}
\end{table}


\baselineskip 18pt
\begin{thebibliography}{99}

\bibitem{e1} D. Bisello et al., (DM2 Collaboration), Phys. Lett. B
{\bf 179}, 294 (1986)
\bibitem{e2} D. Bisello et al., (DM2 Collaboration), Phys. Lett. B
{\bf 241}, 617 (1990)
\bibitem{e3} Z. Bai et al., (MARK III Collaboration), Phys. Rev.
Lett. {\bf 65}, 1309 (1990)
\bibitem{BES08} M. Ablikim et al., (BES Collaboration), arXiv:0801.3885
\bibitem{x1835} M. Ablikim et al., (BES Collaboration), Phys. Rev. Lett. {\bf
95}, 262001 (2005)
\bibitem{BES17602}M. Ablikim et al., (BES Collaboration), Phys. Rev.
D {\bf 73}, 112007 (2006)
\bibitem{pdg2006} W.-M. Yao et al., J. Phys. G {\bf 33}, 1
(2006)
\bibitem{eta1835} De-Min Li, Bing Ma, arXiv:0801.4821
\bibitem{cqm} R. Ricken, M. Koll, D. Merten, B. C. Metsch, H. R.
Petry, Eur. Phys. J. A {\bf 9}, 221 (2000)
\bibitem{regge} A. V. Anisovich, V. V. Anisovich, A. V. Sarantsev,
Phys. Rev. D {\bf 62}, 051502(R) (2000)
\bibitem{3p0rev1} A. Le Yaouanc, L. Oliver, O. Pene, J-C. Raynal,
Phys. Rev. D {\bf 8}, 2223 (1973); Phys. Rev. D {\bf 9}, 1415
(1974); Phys. Rev. D {\bf 11}, 1272 (1975); Phys. Lett. B {\bf
71}, 397 (1977); Phys. Lett. B {\bf 72},57 (1977).
\bibitem{3p0rev2} A. Le Yaouanc, L. Oliver, O. Pene, J-C. Raynal,
Hadron transitons in the quark model ( Gordon and Breach Science
Publishers, New York, 1988)
\bibitem{3p0rev3} W. Roberts and B. Silvestr-Brac, Few-Body Syst.
{\bf 11}, 171 (1992)
\bibitem{3p0rev4} H. G. Blundel, hep-ph/9608473
\bibitem{micu} L. Micu, Nucl. Phys. B {\bf 10}, 521 (1969)
\bibitem{3p00} S. Capstick, N. Isgur, Phys. Rev. D {\bf 34}, 2809
(1986); S. Capstick, W. Roberts, Phys. Rev. D {\bf 49} 4570 (1994)
\bibitem{3p0y} P. Geiger, E. S. Swanson, Phys. Rev. D {\bf 50},
6855 (1994)
\bibitem{3p0x}H.G. Blundell, S. Godfrey, Phys. Rev. D {\bf
53},3700 (1996)
\bibitem{3p0x1} H. G. Blundell, S. Godfrey, B.
Phelps, Phys. Rev. D {\bf 53}, 3712 (1996)
\bibitem{3p0x2} R. Kokoski, N. Isgur, Phys. Rev. D  {\bf 35}, 907
(1987)
\bibitem{3p01} E. S. Ackleh, T. Barnes and E. S. Swanson,
Phys. Rev. D {\bf 54}, 6811 (1996);
\bibitem{3p02} T. Barnes, F. E. Close, P. R. Page and E. S.
Swanson, Phys. Rev. D  {\bf 55}, 4157 (1997)
\bibitem{3p03} T. Barnes, N. Black and P. R. Page, Phys. Rev. D
{\bf 68}, 054014 (2003)
\bibitem{quarkmass}F. E. Close, E. S. Swanson, Phys. Rev. D
{\bf72}, 094004 (2005)
\bibitem{3p04} L. Burakovsky, P. R. Page, Phys. Rev. D {\bf 62}, 014011 (2000); H. Q. Zhou, R. G. Ping, B. S. Zou, Phys. Lett. B
{\bf 611}, 123 (2005); J. Lu, W. Z. Deng, X. L. Chen, S. L. Zhu,
Phys. Rev. D {\bf 73}, 054012 (2006); B. Zhang, X. Liu, W. Z.
Deng, S. L. Zhu, Eur. Phys. J. C {\bf 50}, 617 (2007); F. E.
Close, C. E. Thomas, O. Lakhina, E. S. Swanson, Phys. Lett. B {\bf
647}, 159 (2007); O. Lakhina, E. S. Swanson, Phys. Lett. B {\bf
650}, 159 (2007); C. Chen, X. L. Chen, X. Liu, W. Z. Deng, S. L.
Zhu, Phys. Rev. D {\bf 75}, 094017 (2007); G. J. Ding, M. L. Yan,
Phys. Lett. B {\bf657}, 49 (2007)

\bibitem{mock}C. Hayne and N. Isgur, Phys. Rev. D {\bf 25}, 1944
(1982)
\bibitem{recp} M. Jacob, G. C. Wick, Ann. Phys. {\bf 7}, 404
(1959)

\bibitem{vij} J. Vijande, F. Fernandez, A. Valcarce, J. Phys. G {\bf
31}, 481 (2005)
\bibitem{flavorfun} S. Godfrey, N. Isgur, Phys. Rev. D  {\bf 32},
189 (1985)
\bibitem{mpla} De-Min Li, Bing Ma, Xue-Chao Feng, Hong Yu, Mod. Phys. Lett. A {\bf 20}, 2497
(2005)

\end{thebibliography}
\end{document}